\documentclass[aps,twocolumn,letterpaper, nofootinbib, floatfix, superscriptaddress,amsmath,amssymb]{revtex4-2}
\usepackage{amsmath,amssymb, graphicx}
\usepackage{bm,times, color}
\usepackage[utf8]{inputenc}
\usepackage[colorlinks,breaklinks]{hyperref}
\usepackage[svgnames]{xcolor}
\hypersetup{linkcolor=DarkBlue, citecolor=DarkBlue, filecolor=black, urlcolor=DarkBlue}
\usepackage{comment, mathrsfs,slashed}
\usepackage{amsfonts}
\usepackage{bbm}
\usepackage[normalem]{ulem}
\def \p{\partial}
\def \dag{\dagger}
\def \mb{\mathbf}

\def \lan{\langle}
\def \ran{\rangle}

    \makeatletter
\def\@fnsymbol#1{\ensuremath{\ifcase#1\or \dagger\or \ddagger\or
   \mathsection\or \mathparagraph\or \|\or **\or \dagger\dagger
   \or \ddagger\ddagger \else\@ctrerr\fi}}
    \makeatother

\begin{document}
\title{Electromagnetic fluctuation and collective modes in relativistic bosonic superfluid in mixed dimensions}
\author{Wei-Han Hsiao}
\altaffiliation{Independent Researcher, Chicago, Illinois, USA}
\noaffiliation{}
\date{\today}

\begin{abstract}
In Gaussian approximation, we investigate the marginal electromagnetic fluctuation in models of charged relativistic bosonic superfluids in three and two spatial dimensions at zero temperature. The electromagnetism is modeled by the ordinary Maxwell term and the non-local pseudo-electrodynamics action in these dimensions respectively. We explore the collective excitations in these systems by integrating the superfluid velocity fields. We unveil that different collectives mode dispersions are results of the competition between different characteristic scales of speed and that between short-ranged and long-ranged interactions. In (3+1) dimensions, we derive the roton mode reminiscent of what was discovered in the context of the free relativistic Bose-Einstein condensate as a generalization of the Higgs mode and determine the necessary and sufficient condition for the roton to exist. In (2+1) dimensions, besides solving the dispersion relation for the surface plasmon, we prove there cannot be roton-like excitation in this model as opposed to its (3+1) dimensional counterpart, and additionally derive the asymptotic lines of the dispersion in the limits of long wavelength and short distance. These asymptotic dispersions are supplied with alternative perspective using duality.
\end{abstract}

\maketitle
\section{Introduction}
Many-body bosonic systems with spontaneously broken U(1) symmetry is one of the most popular classes of models introduced to physicists in the many-body curriculum for its richness. On one hand, in the non-relativistic regime, this class containes the celebrated Bose-Einstein condensate, where the underlying particles condense and the particle number of the particles in the ground state becomes classical. It describes the plain vanilla superfluid conventionally applied to ${}^4$He, and establishes a role model for the Ginzburg-Landau description of the fermionic superconductor \cite{landau1980course, schrieffer1999theory}. On the other hand, in the relativistic regime there exists the Abelian Higgs model, where the particle field itself acquires a macroscopic value, which in turn showcases the mechanism by which the fundamental particles become massive, and completes the circle of the current standard model.

Despite the abundance of studies concerning the two examples in the above as well as their fermionic or non-Abelian generalizations, most electromagnetic properties are investigated directly within the framework of Ginzburg-Landau theory in the context of superconductivity and generalized by analogue \cite{weinberg1995quantum, Schwartz:2014sze}. This approach is valid for static properties but often fails to capture the dynamical distinction between relativistic and non-relativistic theories, which is usually characterized by the time derivatives.


In this work, we directly explore the electromagnetic fluctuation of the relativistic superfluid modeled using the Abelian Higgs mode interacting with Lorentz covariant Coulomb interactions in 2 and 3 spatial dimensions. More precisely, we consider the following action
\begin{align}
\label{S_full}S = & \int d^{d+1}x \left( D^{\mu}\Psi^{\dag}D_{\mu}\Psi  - V(\Psi^{\dag}\Psi)\right) + S_{{\rm  EM},d}[A_{\mu}]\notag\\
= & S_{\rm M}[\Psi, A_{\mu}]  + S_{{\rm  EM},d}[A_{\mu}],
\end{align}
where $\Psi$ is a charge boson field with covariant derivative $D_{\mu} = \p_{\mu} + iA_{\mu}$. $V$ is a general potential profile and is assumed to be bounded from below with non-vanishing first and second order derivatives. In particular, by a superfluid we mean $V$ and the chemical potential jointly assume a minimum at finite $\Psi^{\dag}\Psi$. $d$ denotes the dimensions of the space and our electromagnetism is governed by 
\begin{subequations}
\begin{align}
& \label{L21} S_{{\rm EM}, 2} = -\frac{1}{2e^2}\int d^{3}x\, d^3x'\, F_{\mu\nu}\frac{1}{\sqrt{\p^2}}F^{\mu\nu}\\
& \label{L31} S_{{\rm EM}, 3} =  -\frac{1}{4e^2}\int d^4x\, F_{\mu\nu}F^{\mu\nu}, 
\end{align}
\end{subequations}
For $d = 3$, this model describes the relativistic generalization of the Ginzburg-Landau superconductor at zero temperature. The $d = 2$ variant describes a relativistic superconducting film immersed in the $(3+1)$ dimensional electromagnetism. The nonlocal action~\eqref{L21}, sometimes dubbed the pseudo-electrodynamics (PQED) \cite{MARINO1993551}, is a dimension-reduced version of~\eqref{L31} by integrating out the transverse spatial dimension.  

It is worth commenting that we avoid calling the system of interest here a relativistic Bose-Einstein condensate (RBEC) \cite{PhysRevLett.99.200406, Fagnocchi_2010, kapusta_gale_2006} because a RBEC is technically defined by the number equation to which the critical temperature is the solution. In the presence of anti-bosons, the true Noether charge in the number equation corresponds to the difference of numbers of bosons and anti-bosons rather than the condensate $\Psi^{\dag}\Psi$, and does not require a special potential profile to create a vacuum expectation value. Apart from this difference, in the leading order of perturbation theory, the RBEC does share a lot in common with our model and some results echo. Following this thread, in $d = 3$ our model complements the perturbative study in Ref. \cite{REIS2021136003} by the inclusion of potential profile, and in $d = 2$, generalizes part of the discoveries by Ref.\cite{Marino:2017lic} to the regime of relativity.

We shall apply the Gaussian approximation to the model~\eqref{S_full} by integrating out the superfluid density and phase fluctuations. The validity of the perturbation is controlled by the smallness of the marginal quantity $e^2$. The collective excitations in the resulting electrodynamics will be solved. For $d = 3$, we find, similar to a RBEC \cite{REIS2021136003}, the Higgs mode can become a roton. In terms of our model we further deduce that this feature emerges so long as the strength of the contact interaction $V''$ is sufficiently smaller than $e^2$. As for $d = 2$, the general dispersion is solved numerically. Nonetheless we show the surface plasmon cannot be a roton and determine the asymptotic behavior using the technique of duality. Moreover, the results are presented in a manifestly gauge-invariant fashion and rendered interpretable in terms of the competition between velocity scales: speed of light, speed of sound, and phase velocity, and that between interaction strengths: short-ranged density interaction and long-ranged Coulomb interation.

The rest of the paper is organized as follows. In Sec.\ref{sec2}, we first reparametrize the Abelian Higgs model in Eq.~\eqref{S_full} in terms of the superfluid velocity and the magnitude of the condensate. In the Gaussian approximation, we integrate out the fluctuations of magnitude field, resulting in an effective Lagrangian. The full effective electromagnetic theory and gauge invariance are delivered in Sec.\ref{sec3}. Sec.\ref{sec4} is devoted to completely analyzing the collective spectra of the effective electrodynamics and presenting the main results of the paper, followed by the concluding remarks in Sec.\ref{sec5} discussing some potential applications to other systems.
\section{Derivation of the EFT}\label{sec2}
Let us start with the matter part of action~\eqref{S_full} with $A = 0$ temporarily.
\begin{align}
\label{L_0}\mathscr L = \p^{\mu}\Psi^{\dag}\p_{\mu}\Psi  - V(\Psi^{\dag}\Psi)
\end{align}
Without loss of generality, we can parametrize the complex field as $\Psi = \sqrt{n}\, e^{i\theta},$ rephrasing the Lagrangian in terms of the fields of magnitude and phase. 
\begin{align}
\mathscr L =\frac{1}{4n}\p_{\nu}n\p^{\nu}n + n\p_{\nu}\theta\p^{\nu}\theta - V(n).
\end{align}
To slightly demystify the process of spontaneous symmetry breaking, let us separate the fast oscillating part of the field by extracting the characteristic mass scale, denoted by $\mu$. The time dependence is separated from the field $\Psi  \to e^{-i\mu t}\Psi$, or shifting the phase field by the amount $\theta\to \theta - \mu t$, leading to the Lagrangian 
\begin{align}
\mathscr L =\frac{1}{4n}\p_{\nu}n\p^{\nu}n + n\p_{\nu}\theta\p^{\nu}\theta -2\mu n\p_t\theta -V_{\rm{eff}},
\end{align}
where $V_{\rm{eff}} = V(n) - \mu^2 n$. The value of $\mu$ is determined by the solution to the equation of motion. The saddle point solution can be evaluated by the classical extremum of $V_{\rm eff}.$
\begin{align}
\frac{\p V_{\rm eff}}{\p n}\bigg|_{n = \bar{n}} = \frac{\p V}{\p n}\bigg|_{n = \bar{n}} - \mu^2 = 0\Rightarrow \mu^2 = \p_nV(\bar{n}).
\end{align}
An example is the $\phi$-4 theory with $V(n) = m^2n + g n^2/2$ with $m^2, g>0$. In the symmetric phase, $\bar{n} = 0$, implying $\mu^2 = m^2$, the mass of the original particle. For finite $\bar{n}$, we obtain $\mu^2 = m^2 + g\bar{n}$, and thus $\mu$ represents the relativistic chemical potential. Next we rescale the amplitude field by two times the chemical potential $2\mu n\to n$, giving us a form reminiscent of the non-relativistic theory 
\begin{align}
\label{L_eff}\mathscr L = \frac{1}{8\mu n}\p_{\nu}n\p^{\nu}n + \frac{n}{2\mu}\p_{\nu}\theta\p^{\nu}\theta - n\p_t\theta - V_{\rm{eff}}(n,\mu).
\end{align}
The preceding operation is a less standard way to conceive the formation of the Mexican hat potential by turning on the boson density. It has the advantage of making the non-relativistic correspondence transparent by separating fast oscillating fields. We would like to note that this matter Lagrangian with finite $\mu$ cannot fit a superconductor formed by non-relativistic fermions via the BCS mechanism if the particle-hole symmetry of the underlying microscopic model is imposed \cite{Varma2002, doi:10.1146/annurev-conmatphys-031214-014350}. Technically, the Mexican hat in such a BCS Ginzburg-Landau free energy is induced by Cooper instability. The particle-hole symmetry of the underlying fermion forbids the first-order time derivative term, and thus the coupling $n\p_t\theta$, in the resulting effective free energy.

It is now straightforward to turn $A_{\mu}$ back on by shifting $\p_{\nu}\theta \to D_{\nu}\theta = \p_{\nu}\theta + A_{\nu}.$ Note that $\p_{\mu}n$ remains unchanged since $\Psi^{\dag}\Psi$ is charge neutral. Next we linearize the Lagrangians around the classical saddle point $\bar{n}$. 
\begin{align}
\label{L_G}\mathscr L = \frac{1}{8\mu \bar{n}}\delta n( - \p^2-& 4\mu \bar{n}V''(\bar{n}))\delta n \notag\\
& - \delta nD_t\theta + \frac{\bar{n}}{2\mu}D_{\nu}\theta D^{\nu}\theta .
\end{align}
The second order derivatives of the potential can be replaced with the parameter $c_s^2 = \bar n V''(\bar n)/\mu^2$. It has the unit of speed and could be understood as the relativistic version of the speed of sound. The massive amplitude fluctuation $\delta n$ can be integrated out by solving its equation of motion. In terms of the Fourier component, 
\begin{align}
\delta n(p) = \frac{-\frac{\bar n}{\mu}D_t\theta(p)}{c_s^2 - (p^2/(2\mu))^2}
\end{align}
Plugging these solutions back into Eq.~\eqref{L_G}, we cast the Lagrangian in a conventional form of wave equation 
\begin{align}
\mathscr L = \frac{1}{2}\frac{\bar n}{\mu}\left[ \left( 1+ \frac{1}{c_s^2 -(p/2\mu)^2}\right)|D_t\theta|^2 - |\mb D\theta|^2\right]
\end{align}
with a dispersive speed of wave 
\begin{align}
\label{v_general}\frac{1}{v_s^2(p)} = \left( 1+ \frac{1}{c_s^2 -(p/2\mu)^2}\right).
\end{align}
Let us connect this expression to more familiar examples by taking two subsequent limits. First we consider the small frequency limit $\omega /|\mb p|\ll 1$ where the second order temporal variations of $\delta n$ and $\theta$ are omitted. Eq.~\eqref{v_general} simplifies to 
\begin{align}
\frac{1}{v_s^2(p)}\approx \frac{1}{c_s^2 + (\mb p/(2\mu))^2}.
\end{align}
This velocity gives rise to the Bogoliubov mode in the non-relativistic weakly interacting bosons. In the case where the fluctuation at the scale of healing length is further omitted \cite{sym10040080}, the constant velocity can be identified with the ordinary speed of hydrodynamic sound $v_s^2 = c_s^2$.
\section{Higgs mechanism and gauge invariance}\label{sec3}
We proceed to address the fluctuation of the phase field and put the full effective Lagrangian in a manifestly gauge invariant form. As it is shown in the Sec.\ref{sec2} that the relativistic Lagrangians assume the form 
\begin{align}
\label{L_phase} \mathscr L(p) = \frac{1}{2}\frac{1}{\lambda^{d-1}}\left( \frac{|D_t\theta(p)|^2}{v_s^2(p)} - |\mb D\theta(p)|^2\right), 
\end{align}
which is identical to the effective Lagrangian of the superfluid phase in conventional non-relativistic superfluids or superconductors except for the dispersive speed of sound. The parameter $\lambda$ has the unit of length. The Higgs mechanism amounts to integrating the phase field, yielding an effective Lagrangian of $A_{\mu}$ schematically reading 
\begin{align}
\label{EFT} \mathscr L[A] = \frac{1}{2}A_{\mu}(-p)K^{\mu\nu}(p)A_{\nu}(p).
\end{align}
Gauge invariance requires the Ward identity $p_{\mu}K^{\mu\nu}(p) = 0.$ 


To confirm this property explicitly, we solve the saddle point solution to $\p\mathscr L/\p\theta(-p) = 0$: 
\begin{align}
\theta(p) = \frac{-i}{ \omega^2 - v_s^2\mb p^2}\left( \omega A_t(p) + v_s^2\mb p\cdot\mb A(p)\right).
\end{align}
Plugging this solution back to Lagrangian~\eqref{L_phase} immediately entails  
\begin{subequations}
\begin{align}
\label{K00}& K^{00}(p) = -\lambda^{1-d} \frac{\mb p^2}{\omega^2 - v_s^2\mb p^2}\\
\label{K0i}& K^{0i}(p) = -\lambda^{1-d} \frac{\omega p^i}{\omega^2 - v_s^2\mb p^2}\\
\label{Kij}& K^{ij} = \lambda^{1-d}\left[ \frac{p^ip^j}{\mb p^2 - \omega^2/v_s^2} - \delta^{ij}\right].
\end{align}
\end{subequations}
It is then trivial to verify each component of the Ward's identity  
\begin{align*}
\omega K^{00} + p_i K^{i0} \propto \left[ -\omega\mb p^2 + p_i (-\omega p^i)\right] = 0.
\end{align*}
\begin{align*}
& \omega K^{0j} + p_i K^{ij}\propto \left[ \frac{-\omega^2 p^j}{\omega^2 - v_s^2\mb p^2} + \frac{v_s^2\mb p^2p^j}{\omega^2 - v_s^2\mb p^2}-p_j\right] = 0.
\end{align*}
An alternative way to ensure gauge invariance moving forward is to express the Lagrangian directly in terms of the gauge invariant fields $E_i(p) = -i\omega A_i - iq^i A_0$ and $B_k = i\epsilon_{ijk}q_iA_j$, and absorb the dynamical properties into dielectric functions. This way the effective Lagrangian assumes the following form
\begin{align}
\label{inducedEM}\mathscr L = \frac{1}{2}\delta\epsilon(p)|\mb E(p)|^2 - \frac{1}{2}\delta\mu^{-1}(p)|\mb B(p)|^2
\end{align}
with 
\begin{subequations}
\begin{align}
& \delta\epsilon(p) = -\lambda^{1-d}\frac{1}{\omega^2 - v_s^2\mb p^2}\\
& \delta\mu^{-1}(p) = -\lambda^{1-d}\frac{v_s^2}{\omega^2 - v_s^2\mb p^2}
\end{align}
\end{subequations}
being the superfluid induced susceptibilities. In the rest of the paper we shall analyze the propagating waves determined by Eq.~\eqref{inducedEM} together with Eq.~\eqref{L31} and Eq.~\eqref{L21}.
\section{Collective modes in 2 and 3-dimensional spaces}\label{sec4}
Following the preceding steps, let us now append the electromagnetic fluctuations to Eq.~\eqref{inducedEM}. The exercise amounts to deriving the zeros in generalized Maxwell's equations.
\begin{subequations}
\begin{align}
& (\epsilon_d + \delta\epsilon)i \mb q\cdot\mb E = 0\\
& (\epsilon_d + \delta\epsilon) i \omega\mb E + (\mu^{-1}_d + \delta\mu^{-1})i \mb q\times\mb B = 0.
\end{align}
\end{subequations}
Identical to the elementary electromagnetism, the longitudinal mode is given by the zeros of the kernel of the Gauss law, whereas the transverse mode is given by the generalized Ampere and Faraday's laws $i \mb q\times\mb E = i\omega\mb B$, which implies
\begin{align}
[(\epsilon_d + \delta\epsilon)\omega^2 - (\mu^{-1}_d + \delta\mu^{-1})\mb p^2]\mb B = 0.
\end{align}
\subsection{d = 3}
In $d = 3$, $\epsilon_3 = \mu_3^{-1} = e^{-2}$. The transverse mode is given by the homogeneous equation
\begin{align}
[e^{-2}(\omega^2 - \mb p^2) - \lambda^{-2}]\mb B = 0.
\end{align}
By defining the {\it plasmon mass} $m_p^2= e^2\lambda^{-2}$, the dispersion relation is simply 
\begin{align}
\omega_T^2 = m_p^2 + \mb p^2.
\end{align}
The relativistic effect comes in through the relativistic chemical potential $\mu = m + \mu_{\rm{nr}}$ but it does not modify the dispersion of the electromagnetic wave. As such relativistic effect affects the numerical value of the penetration length but is not going to modify the decay profile caused by the Meissner effect. Besides, we can also recognize the length scale $\lambda$ roughly corresponds to the London penetration length in the model.

To derive the longitudinal mode $\mb p\cdot\mb E\neq 0$, we look at the Gauss law 
\begin{align}
\left[ \frac{1}{e^2} - \lambda^{-2}\frac{1}{\omega^2 - v_s^2\mb p^2}\right] i \mb p\cdot\mb E = 0,
\end{align}
which yields the dispersion relation 
\begin{align}
\omega^2 = m_p^2 + \frac{\mb p^2}{1 + \frac{1}{c_s^2 - (p/2\mu)^2}}.
\end{align}
To solve it, it is convenient to introduce the dimensionless plasmon mass, frequency and wavenumber $(M, \Omega, \mb P) =\frac{1}{2\mu} (m_p, \omega, \mb p)$. Unwinding the quadratic equation, the roots are given by 
\begin{align}
\label{QED4sol}& 4\Omega^2 + (c_s^2 - M^2)^2 - (2M)^2\notag\\
 =& (1\pm\sqrt{(1 + c_s^2 - M^2)^2 + 4\mb P^2})^2.
\end{align}
The limiting cases of these modes can be read off. On one hand, at short length scale, $\mb P\gg 1$, both modes have the same asymptote $\omega \approx |\mb p|$. On the other hand, in the long-wavelength limit $\mb P^2 = 0$, we can easily solve two gap modes 
\begin{subequations}
\begin{align}
\label{gap0}& \omega^2 = m_p^2\\
\label{gap1}& \omega^2 = 4\mu^2 (1 + c_s^2).
\end{align}  
\end{subequations}
We note that these two gaps are of different nature. Mode~\eqref{gap0} depends on the electromagnetic coupling $e^2$ and its gap closes as the electromagnetism is turned off. This mode is created by Higgs mechanism and corresponds to the relativistic version of the plasmon oscillation. Mode~\eqref{gap1}, on the other hand, is independent of the strength of $e^2$, depending upon the chemical potential and sound velocity of the original superfluid. This mode originates from the {\it pair production} mechanism of the original boson field. As we can confirm by draining the particle density such that $\mu \to m$ and $c_s^2 = 0$, in the limit of which $\omega \to 2m$. 

\begin{figure}
 \includegraphics[width=0.9\linewidth]{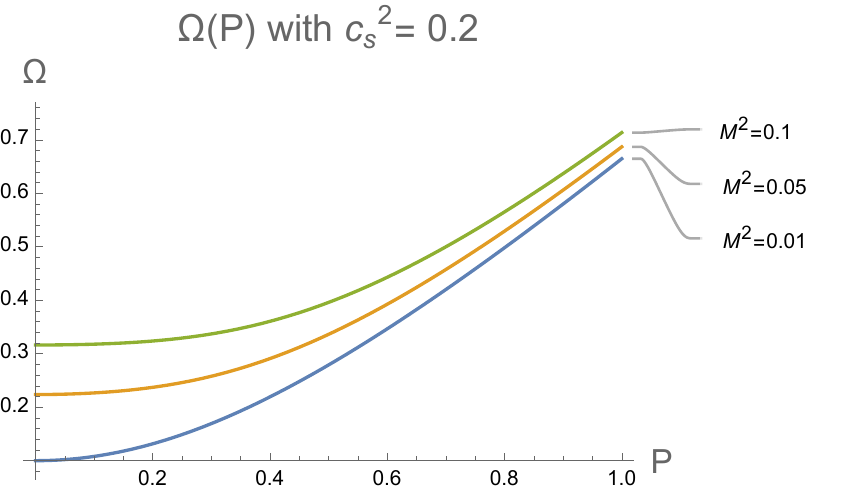}
 \includegraphics[width=0.9\linewidth]{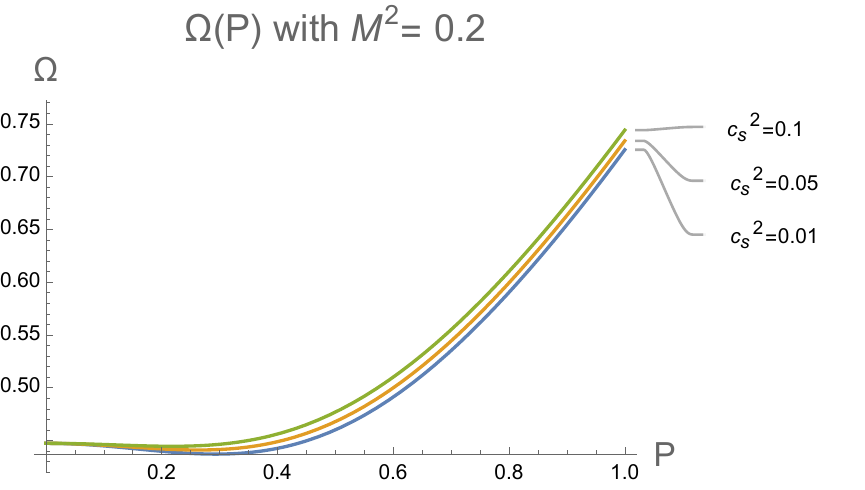}
 \caption{Illustrations of solution~\eqref{QED4sol} with various values of $c_s^2$ and $M^2$. The lower panel clearly shows the existence of local minimum when $M^2>c_s^2$.}
 \label{fig0}
\end{figure}

The general dispersion of these modes depends on the magnitudes of $c_s^2$ and $M^2$, or the magnitudes of the short-ranged density-density interaction term $V''(\bar n)$ and the long-ranged electromagnetic coupling $e^2$ if we trace back to the original field theory model. We plot several examples with various tuples of parameters in Fig.\ref{fig0}.

The geometric properties of the dispersions become more transparent in terms of their slopes.
\begin{align*}
\frac{\p \Omega}{\p \mb P^2} = \frac{1}{2\Omega}\frac{\p \Omega^2}{\p\mb P^2}\propto \pm 1 + \sqrt{(1-M^2 + c_s^2)^2 + 4\mb P^2}
\end{align*}
As this formula entails, for the positive branch, the slope is positive definite and it simply interpolates the pair-production gap~\eqref{gap1} to the photonic asymptote. For the negative plasmon branch, the story is not too different in the regime $c_s^2\gg M^2$. Nevertheless, it is clear that as $M^2 > c_s^2$, or equivalently
\begin{align}
e^2>2 V''(\bar n),
\end{align}
the following inequality holds
\begin{align*}
-1 + \sqrt{(1-M^2 + c_s^2)^2 + 4\mb P^2} < 0
\end{align*}
for a range of $\mb P^2$. This leads to a {\it roton}-like dispersion in the bosonic superfluid with the gap 
\begin{align}
\Omega^2 = M^2\left( 1 - \frac{1}{4M^2}(M^2-c_s^2)\right) < M^2.
\end{align}
We emphasize that this is a pure relativistic effect because the competition between $M^2$ and $c_s^2$ is induced via the second time derivative of $\theta$ field. Otherwise, $M^2$ only determines the plasmonic gap, never affecting the dispersion. To be precise, we can repeat the steps in the previous section without the term $|D_t\theta|^2$ and $(\p_{t}n)^2$. With good accuracy the only collective mode is 
\begin{align}
\omega^2 = m_p^2  + {\mb p^2}(c_s^2 + (\mb p/2m)^2).
\end{align}
Its derivative is obviously independent of $m_p^2$ and positive definite, being identical to that of the the Bogoliubov quasiparticle
\begin{align*}
\frac{\p\omega^2}{\p \mb p^2} = c_s^2 + \left(\frac{\mb p}{m}\right)^2.
\end{align*}
As such one cannot infer this {\it rotonic} feature from non-relativistic perspective. This regime is also not feasible for conventional fermionic superfluid or the BCS model because the speed of sound in those systems is of the order of the Fermi velocity $v_F$, which usually predominates the plasma frequency. Moreover, the above argument merely assumes the relative hierarchy $M^2>c_s^2$ rather than the absolute magnitude of either quantity. So long as the inequality holds, the slope is negative for a finite range of $\mb P^2$. Therefore, the signature should survive even in the perturbative regime where $e^2$ and $V''(\bar n)$ are numerically small.

To the best of our knowledge, a {\it roton mode} of this kind was recently discovered in a RBEC using diagrammatic approach in Ref.\cite{REIS2021136003}. It is nonetheless worth pointing out the condensate in that work is generated by a heuristic substitution of the quasiparticle distribution rather than a potential profile encoded in the Lagrangian. Our approach makes it clear the condition for these rotons to emerge by investigating the competition between short-ranged and electromagnetic interactions.

The plasmon branch naturally contains the known collective excitations in more conventional regimes. In the non-relativistic and long-wavelength limit, $\mu\to m$ and healing length correction drops out,
\begin{align}
\omega^2 = \frac{\bar n e^2}{m} + c_s^2 \mb p^2.
\end{align}
\subsection{d=2}\label{sec42}
In 2 spatial dimensions, the relativistic Coulomb interaction is given by the non-local Maxwell term. $\epsilon_2 =\mu_2^{-1} = \frac{2}{e^2\sqrt{-q^2}}$, leading to the equation for transverse mode
\begin{align}
\left(\frac{\omega^2 - \mb p^2}{\sqrt{\mb p^2 - \omega^2}} - \frac{e^2}{2\lambda}\right) B( p) = 0.
\end{align}
This equation does not have any propagating wave solution. This conclusion is aligned with the model for thin superconducting film \cite{Marino:2017lic} because as we pointed out the spacetime dispersion of the Goldstone mode does not induce correction to the transverse component of U(1) gauge field. 

Moving onto the longitudinal mode given by solving 
\begin{align}
\left( \frac{2}{e^2}\frac{1}{\sqrt{-p^2}} - \frac{\lambda^{-1}}{\omega^2 - v_s^2\mb p^2}\right) i \mb p\cdot\mb E = 0.
\end{align}
Again in terms of the dimensionless quantities $(\Omega, \mb P, M') = \frac{1}{2\mu}(\omega, \mb p, e^2/(2\lambda))$, the above is cast into the following equation 
\begin{align}
\label{toSolve}(\Omega^2- \mb P^2) + \frac{\mb P^2}{1 + c_s^2 - (\Omega^2 - \mb P^2)} = M'\sqrt{\mb P^2- \Omega^2}.
\end{align}
At $\mb P = 0$, we still have the pair-breaking mode $\Omega^2 = 1+c_s^2$ and the plasmon $\Omega^2 = 0.$ The former nonetheless cannot propagate as imaginary part develops at finite $\mb P$ for any finite electromagnetic interaction $e^2\neq 0$. This can be shown directly by plugging the ansatz $\Omega^2 = 1 + c_s^2 + \frak a\mb P^2$ into Eq.~\eqref{toSolve} and solve $\frak a$ assuming $0<\mb P^2\ll 1$. It can be found $\mathrm {Im}[\frak a] \propto M'\neq 0$. 
\begin{figure}
 \includegraphics[width=0.9\linewidth]{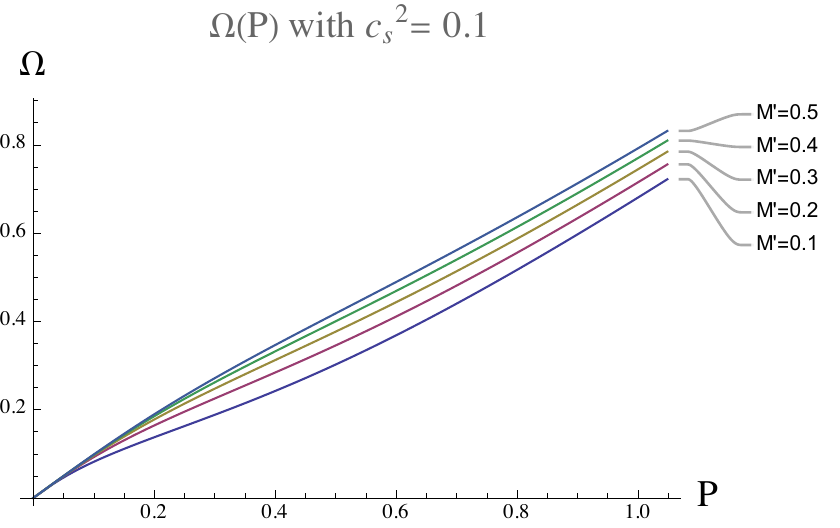}
 \includegraphics[width=0.9\linewidth]{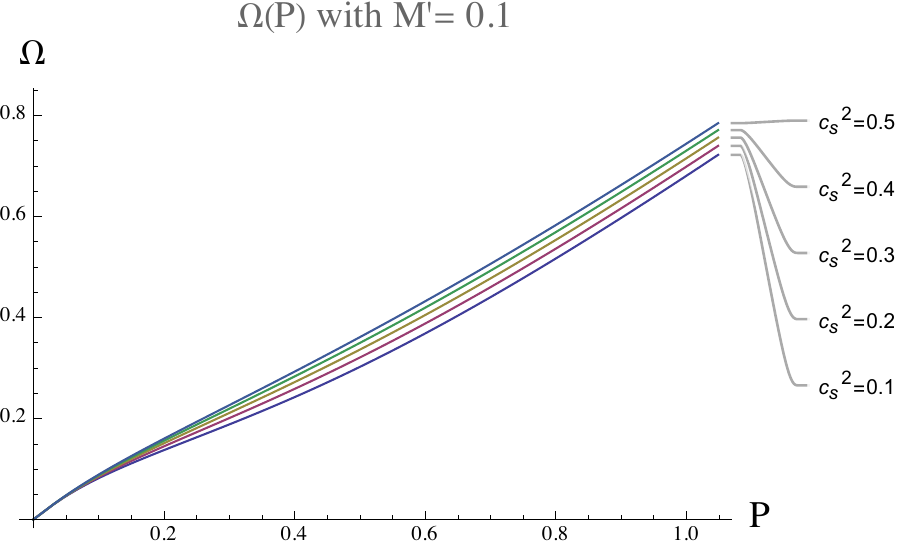}
 \caption{The numerical solutions to Eq.~\eqref{toSolve} with various values of $M'$ and $c_s^2$. In particular, the upper and lower graphs fix the value of $c_s^2$ and $M'$ respectively in order to visualize the variation of the dispersion relations as the other parameter varies.}
 \label{fig1}
\end{figure}

For general $(c_s^2, M')$, the dispersion for propagating plasmon mode is solved numerically as shown in Fig.\ref{fig1}. Without solving the equation exactly, we can deduce some asymptotic behaviors in the limits of long wavelength $|\mb P^2|\ll 1$ and short distance $|\mb P|\gg 1$. The wavelike solutions $\omega  = v|\mb p|$ in these regimes both have $v\sim 1$, i.e., the ordinary speed of light. In between these extrema the group velocity first decreases, deviating away from ordinary electromagnetic wave, then increasing to catch up with the light asymptotically. We could derive an approximation for the dispersion in this nonlinearly dispersive regime for small $|\mb P|$ if there exists a separation of scales $c_s^2 \ll 1$. Given the smallness of $c_s^2$ as well as the asymptotic feature at small $|\mb P|$, Eq.~\eqref{toSolve} reduces to 
\begin{align}
\Omega^2 \approx M'\sqrt{\mb P^2 - \Omega^2},
\end{align}
which yields the solution 
\begin{align}
\Omega^2 = \frac{M'}{2}(\sqrt{(M')^2 + 4\mb P^2} - M').
\end{align}
Thus the dispersion starts along with $\omega = |\mb p|$ from $|\mb p| = 0$ and becomes concave following roughly the shape $\Omega \approx ((M')^2\mb P^2)^{1/4}$.

We would like to point out the asymptotic behavior at large $|\mb P|$ differs than the result derived from non-relativistic superconducting film subject to the same electromagnetic fluctuation studied in Ref.\cite{Marino:2017lic}. The discrepancy is caused by the fact that the model in Ref.\cite{Marino:2017lic} omits both relativistic effect and fluctuations at the scale of healing length. In particular, omitting the latter completely eliminates the spatial dispersion of the speed of sound $v_s^2$, so $c_s^2\mb p^2$ term in $v_s^2\mb p^2$ predominates at large $\mb p$. To deduce this signature in our model, let us consider $c_s^2<1$ and zoom into the vicinity of $\mb P= 0$. At a long enough wavelength we have $\Omega^2 - \mb P^2 \ll c_s^2$, validating the expansion 
\begin{align*}
\frac{\mb P^2}{1 + c_s^2 + (\Omega^2 - \mb P^2)}\approx (1-c_s^2)\mb P^2.
\end{align*}
The lefthand side in turn becomes $\Omega^2 - c_s^2\mb P^2$. 

The asymptotic behaviors of the dispersion relations can be further elaborated  in (2+1) dimensions by the method of duality. It is done by directly Legendre transforming the Gaussian Lagrangian~\eqref{L_phase}. With a Hubbard-Stratonovich transformation, we first introduce a vector field $J^{\mu}$
\begin{align}
\mathscr L = -\frac{\lambda}{2}(v_s^2|J^0(p)|^2 - |J^i(p)|^2) + J^{\mu}(-p)D_{\mu}\theta(p).
\end{align}
In this formulation the $\theta$ field becomes a Lagrange multiplier that enforces the constraint $\p_{\mu}J^{\mu} = 0$. In (2+1) dimensions, we solve it by introducing an auxiliary gauge field $a_{\mu}$ with $J^{\mu} = \frac{1}{2\pi}\epsilon^{\mu\nu\lambda}\p_{\nu}a_{\lambda}$.The above then becomes a QED3 coupled to a  PQED term.
\begin{align}
\mathscr L_{\rm Dual} = & \frac{\lambda}{8\pi^2} (\mb e^2 - v_s^2 b^2) \notag\\
& + \frac{1}{2\pi} \epsilon^{\mu\nu\lambda}A_{\mu}\p_{\nu}a_{\lambda} -\frac{1}{2e^2}F_{\mu\nu}\frac{1}{\sqrt{\p^2}}F^{\mu\nu}
\end{align}
with a dispersive speed $v_s^2$. The asymptotic behaviors can now be read off more easily using this Lagrangian by counting the spacetime derivatives acting on the Maxwell terms of $(\mb e, b)$ and $(\mb E, B)$. At long wavelength and small frequency, the $1/\sqrt{-p^2}$ is dominant and the model reduces to the charge-free PQED 
\begin{align*}
{\mathscr L_{\rm Dual}} \to -\frac{1}{2e^2}F_{\mu\nu}\frac{1}{\sqrt{\p^2}}F^{\mu\nu}.
\end{align*}
The dispersion in this model is trivially $\omega = |\mb p|$. In the opposite limit of short distance, the dynamics of the dual Lagrangian is mostly attributed to 
\begin{align*}
\mathscr L_{\rm Dual} \to  \frac{\lambda}{8\pi^2} (\mb e^2 - v_s^2 b^2).
\end{align*}
implying a dispersion 
\begin{align}
\omega^2 = v^2(p)\mb p^2.
\end{align}
Solving this equation yields 
\begin{align}
& \Omega^2 =  \mb P^2 + \frac{1}{2}\left( 1 - \sqrt{(1+c_s^2)^2 + 4\mb P^2}\right)\notag\\
\Rightarrow & \omega \approx |\mb p| - \mu.
\end{align}
This gives us an estimate for the second asymptote. In the approximation where $v_s^2(\mb p) = c_s^2$ the result in Ref.\cite{Marino:2017lic} is also consistently derived. 

To conclude this section, we would like to prove, unlike $d=3$, there cannot exist rotonic propagating mode in the model. First we acknowledge that $\mb P^2-\Omega^2 > 0$ in order for the mode to be propagating. It suffice to show under no circumstances $\p\Omega^2/\p\mb P^2 = 0$ for any solution to Eq.~\eqref{toSolve} because a roton is usually defined by a local minimum of dispersion curve at a finite wavenumber. Taking the derivative with respect to $\mb P^2$ on both sides of Eq.~\eqref{toSolve}, we conclude
\begin{align}
&\left( 1- \frac{\p\Omega^2}{\p \mb P^2}\right) \left[ 1 + \frac{\mb P^2}{(1+c_s^2 + \mb P^2-\Omega^2))^2} + \frac{M'}{2\sqrt{\mb P^2-\Omega^2}}\right]\notag\\
 =& \frac{1}{1 + c_s^2 + \mb P^2 - \Omega^2}.
\end{align}
If $\p\Omega^2/\p\mb P^2 = 0$ at some $\mb P$ given $M'$ and $c_s^2$, the terms in the bracket on the lefthand side has to match the righthand side, or 
\begin{align}
\label{necessary_condition}1 = 1 + c_s^2 &+ (\mb P^2  -\Omega^2)  + \frac{\mb P^2}{1 + c_s^2 + \mb P^2 - \Omega^2}\notag\\
 & + \frac{M'}{2\sqrt{\mb P^2 - \Omega^2}}(1 + c_s^2 + \mb P^2 - \Omega^2).
\end{align}
Nonetheless, the righthand side of Eq.~\eqref{necessary_condition} is greater than one, implying the slope cannot vanish on any occasion. Consequently, there is no rotonic longitudinal mode in this model.

\section{Concluding remark}\label{sec5}
At the level of Gaussian approximation, we have shown in a gauge invariant fashion that the Abelian Higgs model exhibits different electromagnetic properties when accounting for the relativistic time dependence of the superfluid density and phase fields. The inclusion introduces the competition between multiple {\it scales of speeds}, the speed of light $c=1$, the speed of sound $c_s$, and the dimensionless wavenumber. It also induces extra competition between the short-range contact and long-range Coulomb interactions. These competitions lead to new features distinct from the ordinary non-relativistic bulk and surface plasmons such as the roton-like excitation and the asymptotic behavior at short distance. For spatial dimensions $d=2$, we elaborated different limits where the dispersions have interpretable analytic forms, and provided alternative point of views using particle-vortex dualities. On top of these, we supply a no-go argument for the roton-like excitation based on investigating the curvature of the dispersion. 

We expect these results could apply to a generic class of charged relativistic superfluids, and note that the potential candidates overlap with ones of RBEC because of the fundamental resemblance. To the best of our knowledge, on one hand, generic bosonic superfluids in labs are charge neutral as the long-range Coulomb potential poses a technical challenge. In nature, on the other hand, results in $d=3$ can possibly realize in the deep interior of a neutron star, where various charged bosons such as mesons and pions may condense \cite{KOWALENKO1985109, PhysRevD.86.064011, kolo2018}.

Moving forward, for $d=3$ it is worth exploring the roton regime at finite temperature so as to fill the blank chapter between our result, $(T= 0, V''\neq 0)$ and one from RBEC, $(T\neq 0, V'' = 0)$ \cite{REIS2021136003}. This regime is relevant for empirical observations enumerated in the above.

As for the mixed-dimensional system ($d = 2$) the discoveries in the main text could be applied to other strongly correlated phases by virtue of relativistic particle-vortex duality \cite{SEIBERG2016395, PhysRevX.6.031043} and the strong-weak dual structure from the mixed-dimensional QED \cite{PhysRevB.100.235150, PhysRevD.104.125006}. For instance, suppose we specify an explicit $V = -r(\Psi^{\dag}\Psi) + \frac{g}{2}(\Psi^{\dag}\Psi)^2$, $r, g > 0$, in Eq.~\eqref{S_full}. By the relativistic particle-vortex duality, this action is dual to another boson model \cite{ PESKIN1978122, PhysRevLett.47.1556, PhysRevB.100.235150, PhysRevD.104.125006} 
\begin{align*}
(D_{\nu}\Phi)^{\dag}D^{\nu}\Phi + r'(\Phi^{\dag}\Phi) + \frac{g'}{2}(\Phi^{\dag}\Phi)^2 + \cdots + S_{{\rm EM}, 2}[A, \tilde e]
\end{align*}
with $\tilde e = (4\pi)/e$ and $r', g' >0$. The potential profile implies a mean-field phase of Mott insulator subject to a background magnetic field proportional to the particle density of $\Psi$. Given $\bar n \neq 0$, the charge density 
\begin{align*}
\lan \rho_{\Psi}\ran = -\lan \frac{\delta S_{\rm M}}{\delta A_0}\ran = \lan n - \frac{n}{\mu}D_t\theta\ran \approx \bar n.
\end{align*}
Note that the shift $\Psi \to e^{-i\mu t}\Psi$ already separate the chemical potential from the full $A_0$. As a consequence, the model studied in Sec.\ref{sec42} is dual to a Mott insulator in a Landau level subject to strong electromagnetic fluctuation by this back-of-the-envelope derivation. The detailed account for the relativistic particle-vortex duality is nonetheless beyond the scope of this paper and we defer it to dedicated future investigations. 
\begin{acknowledgements}
The author thanks Umang Mehta, Hart Goldman, Hsiang-Chih Hwang, and Ti-Lin Chou for intriguing discussion, and Yu-Ping Lin for comments on the early version of the paper. Besides, the author is extremely grateful for the encouragement from Shih-An Wang and professor Chiao-Hsuan Wang.
\end{acknowledgements}
\bibliography{citation}
\end{document}